\documentclass{aip-cp}

\usepackage[numbers]{natbib}
\usepackage{rotating}
\usepackage{graphicx}

\begin{document}

\title{NectarCAM, a camera for the medium sized telescopes of the Cherenkov Telescope Array}

\author[aff1]{J-F. Glicenstein\corref{cor1}}
\author[aff1]{M.Shayduk}
\author[aff2]{the NectarCAM Collaboration}
\author[aff3]{the CTA Consortium}
\affil[aff1]{CEA/DRF/IRFU, CEA-Saclay, 91191 Gif-sur-Yvette, France}
\corresp[cor1]{Corresponding author: glicens@cea.fr}
\affil[aff2]{
APC (Paris), CIEMAT (Madrid), CPPM (Marseille), DESY (Zeuthen), ICC-UB(Barcelona), IFAE-BIST(Barcelona), IPAG(Grenoble), IPN(Orsay), IRAP(Toulouse), IRFU(Saclay), LAPP(Annecy), LLR(Palaiseau), LPNHE(Paris), LUPM(Montpellier), UCM-GAE(Madrid)
}
\affil[aff3]{see www.cta-observatory.org for full author \& affiliation list}
\maketitle

\begin{abstract}
 NectarCAM is a camera proposed for the medium-sized telescopes of the Cherenkov Telescope Array (CTA) which covers the core energy range of  ~100 GeV to ~30 TeV.  It has a modular design and is based on the NECTAr chip, at the heart of which is a GHz sampling Switched Capacitor Array and 12-bit Analog to Digital converter. The camera will be equipped with 265 7-photomultiplier modules, covering a field of view  of 8 degrees. Each module includes photomultiplier bases, high voltage supply, pre-amplifier, trigger, readout and Ethernet transceiver.  The recorded events last between a few nanoseconds and tens of nanoseconds. The expected  performance of the camera are discussed. Prototypes of NectarCAM components have been built to validate the design. Preliminary results of a 19-module mini-camera are presented, as well as future plans for building and testing a full size camera. 
\end{abstract}

\section{Concept}
\begin{table}[h]
\caption{
Main specifications of NectarCAM}
\label{tab:properties}
\tabcolsep7pt\begin{tabular}{ll}
\hline
  \tch{1}{c}{b}{Parameter}  & \tch{1}{c}{b}{Value}  \\
\hline
Camera field of view & 8 degrees \\
Number of photodetectors (pixels)   & 1855\\
Weight & 1930 kg \\
Number of modules & 265 \\
Physical size & 2.8 m x 2.9 m x 1.15 m \\
Optical transmission & $\sim 18$ \% (measurement+estimation)\\
Analogue bandwidth & $> 250$ MHz (measured) \\
Integration window & 1--60 ns \\
Trigger threshold & 80 GeV \\
Sampling frequency & 0.5--2 GHz (nominal 1 GHz) \\
Dynamic range ($<5$\% non linearity) & 0.5-2000 photo-electrons (measured) \\
Charge resolution & 32\% (single photoelectron), $\sim 2$ \% (2000 p.e) \\
Time resolution & $<1$ ns ($>$ 5 p.e) \\
Dead time & $< 3$ \% at 4.5 kHz trigger rate \\
Power consumption (embedded part) & 7.7 kW (peak 8 kW) \\
Power consumption (on-ground part) & 3 kW \\
\hline
\end{tabular}
\end{table}

NectarCAM is a Cherenkov camera designed for the
medium sized telescopes of CTA. The basic element is the Nectar module, shown in Fig. \ref{fig:module} described later in this paper. The design of NectarCAM is built on the knowledge  gained from that of the H.E.S.S. and MAGIC cameras. However, NectarCAM improves on these cameras in several aspects. The field of view is much larger, and signals of energetic showers can last up to more than 60 ns.  Read-out windows are thus larger than in the H.E.S.S. cameras. The data are sampled at 1 GHz and entire waveforms are recorded. The data rate can be as high as 30 Mbit/s. Data from the 265 Nectar modules are first concentrated over four 10Mbit Ethernet links before being sent to a camera server. The
reliability of NectarCAM is enhanced by protecting the electronics from enviromental effects , by sealing
the camera and by carefully controlling the temperature. Cool air is circulated inside the Nectar modules to keep temperature between 25 and 35 degrees.
The main specifications of NectarCAM are
listed in Table \ref{tab:properties}.

\begin{figure}[h]
  \centerline{\includegraphics[width=12cm]{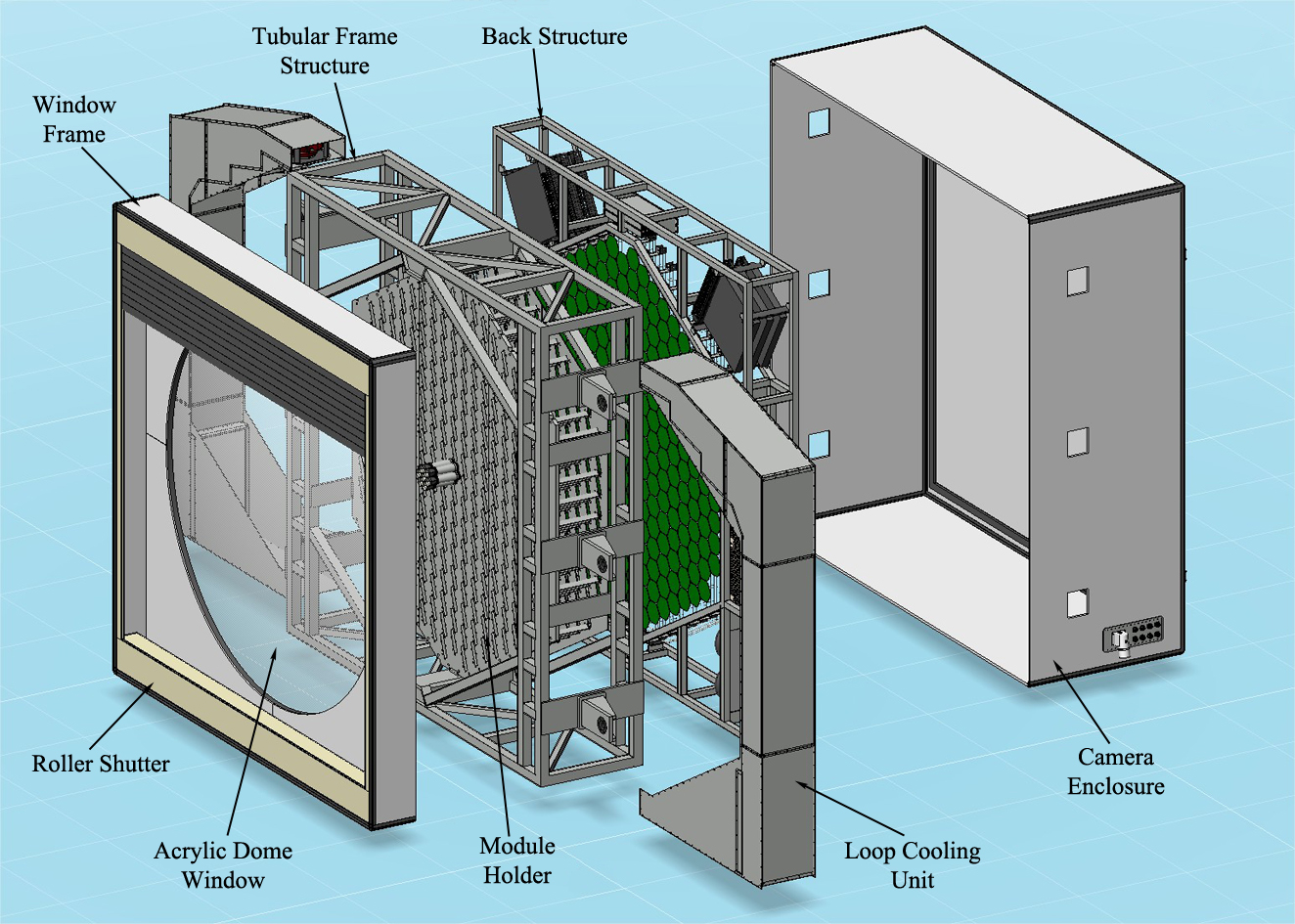}}
  \caption{Exploded view of NectarCAM}
  \label{fig:camera_exploded}
\end{figure}

NectarCAM is built from several major independent assemblies or units (Fig. \ref{fig:camera_exploded}). This architecture provides flexibility to the construction and the integration of the cameras during the production phase.
The front assembly, the camera entrance aperture, includes a customized commercial roller shutter to blind the camera and an acrylic sheet that provides the sealing function. Both are mounted on an aluminum frame structure. The front assembly is hinged to the central structure of the camera to provide easy access the modules. 
The central assembly comprises the primary load bearing component of the camera, a tubular frame structure made of welded aluminum profiles, the module holder that receives all the Nectar modules, and the cooling units of the front-end electronics. The NectarCAM cooling system has been tested on a prototype, the results of tests are described in Reference \cite{Emmanuel}.   
The back assembly holds the trigger and data acquisition subsystems and auxiliary systems of the camera. 
The devices are mounted on a light self-supporting aluminum structure. 
The camera housing is made from large aluminum honeycomb panels. 

 All the individual hardware components of NectarCAM have been prototyped and tested succesfully. 
Many of of these components were designed in common with the LSTCAM camera
which will be installed on the large-sized telescopes of CTA. 
The software architecture has been designed and a large fraction of the components has been prototyped. Finally, a part of the NectarCAM design was driven by interfaces to the MST telescope structure (mechanical interfaces, power and time distribution)  \cite{Gerd} and the CTA control and acquisition system. 
\section{Nectar modules}\label{sec:nectarmodule}

\begin{figure}[h]
  \centerline{\includegraphics[height=6cm]{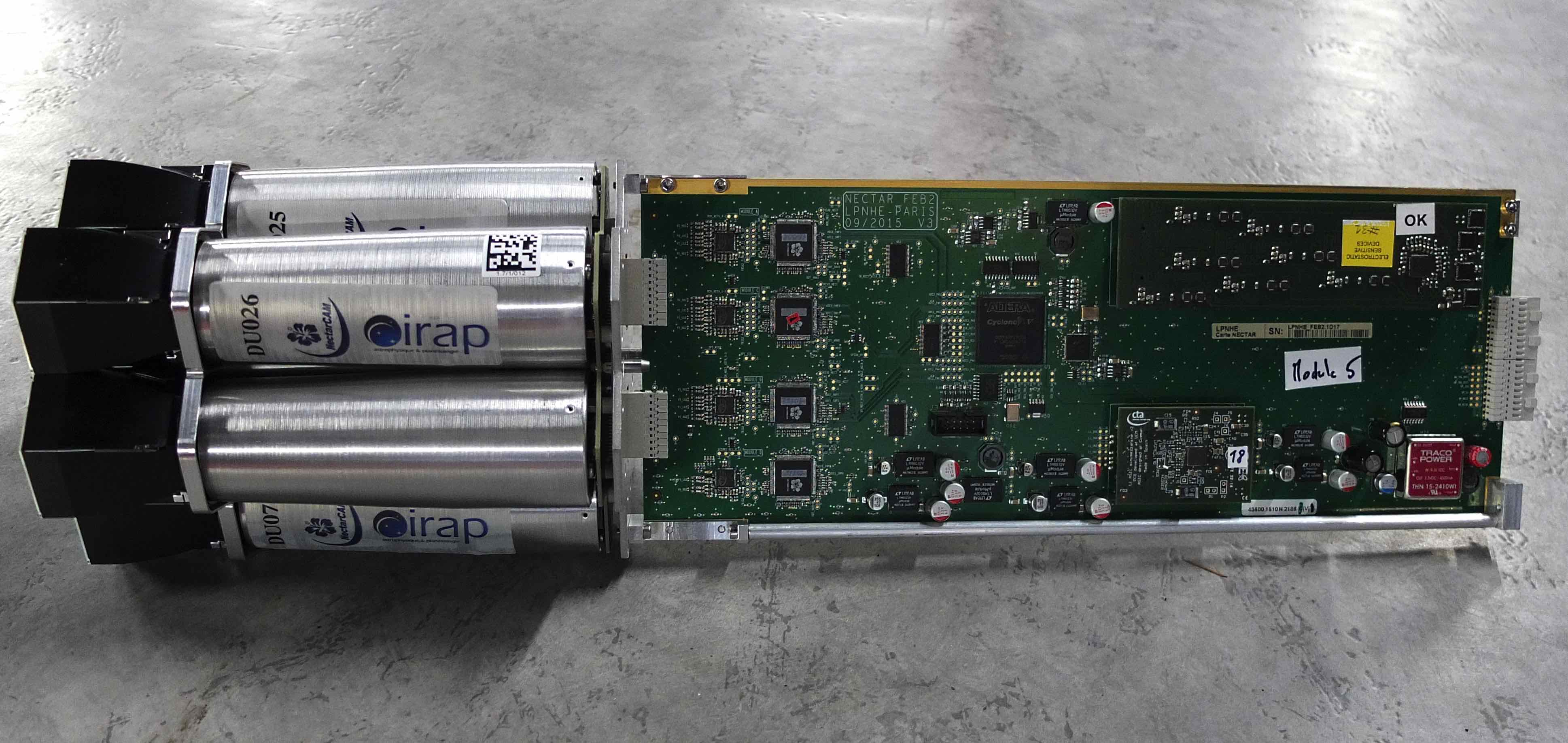}}
  \caption{Nectar module. Light is detected in detector units (left-hand side) and stored in a switched capacitor array on the front end board (right-hand side). After evaluation of local L0 (front-end board level)/L1 (camera level)  trigger conditions, a region of interest of 60 ns is read-out and sent to a camera server.}
  \label{fig:module}
\end{figure}

The basic detection unit of NectarCAM is the Nectar module, shown on Fig. \ref{fig:module}. It is composed
of a focal plane module with 7 {\em detector units} connected to the front-end board by an Interface
Board. It is powered by a 24 V power supply, and receives a pulse-per-second (pps) signal from the CTA time network \cite{2015arXiv150901164O}.
The pps signal is used to synchronize local event counters of modules.   

 Each detector unit includes a R12992-100 7-dynode phototube from Hamamatsu as well as a HVPA
board that contains a Cockroft-Walton
high voltage generator and a signal preamplifier. The signal is
amplified a second time on the front-end board in an ASIC named ACTA \cite{2010JInst...512034G} and divided into two data and one trigger channel.
Two data channels with different gains are needed to accomodate the large dynamic range of 0.5-2000 photoelectrons.
The data channels are sampled at a 1GHz frequency and stored in a switched capacitor array in the
NECTAr chip. The NECTAr chip is described in detail in reference \cite{6154348}. 
When a trigger occurs, a region of interest of up to 60ns is read-out in the two gain channels of every pixel, 
digitized with a 12 bit accuracy by a Wilkinson ADC -- also included  in the NECTAr chip -- and sent to a
camera server through an Ethernet connection. The front end board
also holds local (L0, L1) trigger components and
is connected to a backplane that performs the
camera trigger decision. 

\begin{figure}[h]
  \centerline{\includegraphics[height=6cm]{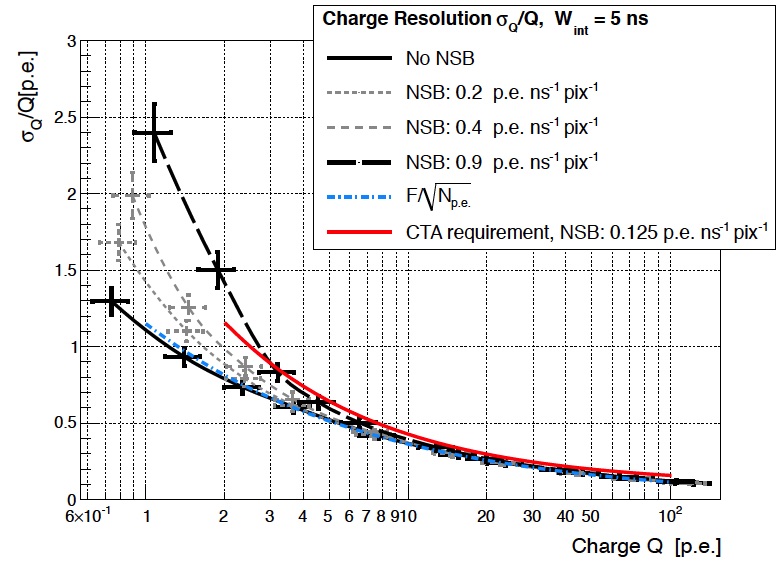}\includegraphics[height=6cm]{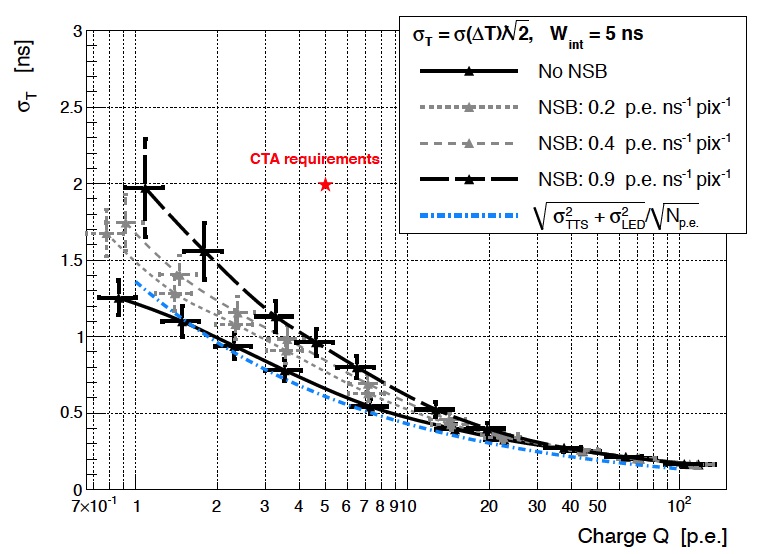}}
  \caption{Preliminary charge (left) and time (right) resolutions obtained with a Nectar module, compared to the CTA requirements. The result is shown for several night sky background levels.
}
  \label{fig:resolutions}
\end{figure}

The ouput of a NectarCAM modules is linear, with less than 5\% non-linearity over the whole dynamic range.  The time and charge
resolutions obtained with a Nectar module are shown in Fig. \ref{fig:resolutions}. A LED providing continuous green light was used to provide different levels of Night Sky Background (NSB). The NSB level was measured with a PIN diode equipped with Winston cones placed close to the Nectar module.

\section{19-module mini-camera}

\begin{figure}[h]
  \centerline{\includegraphics[height=6cm]{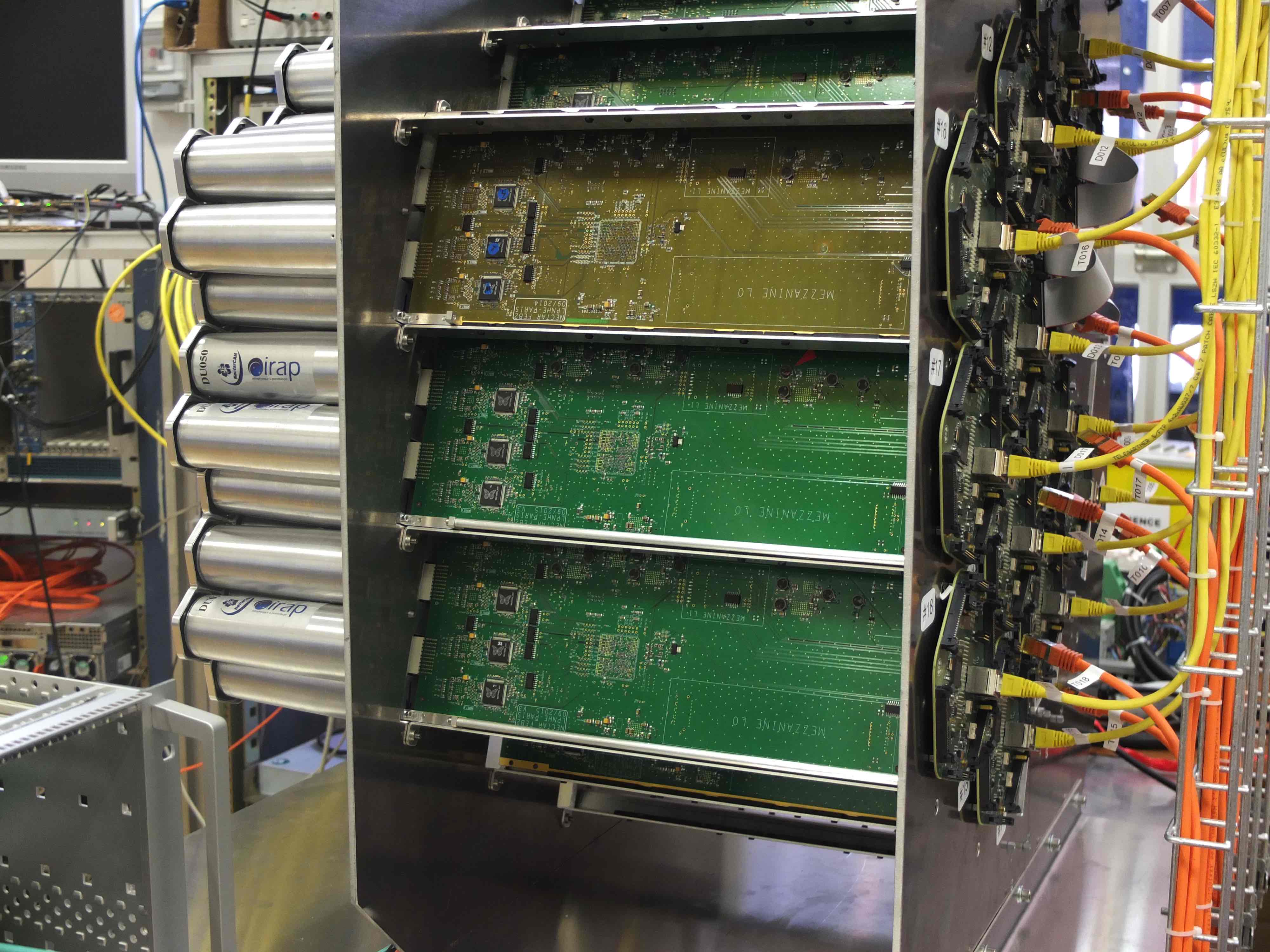}\includegraphics[height=6cm]{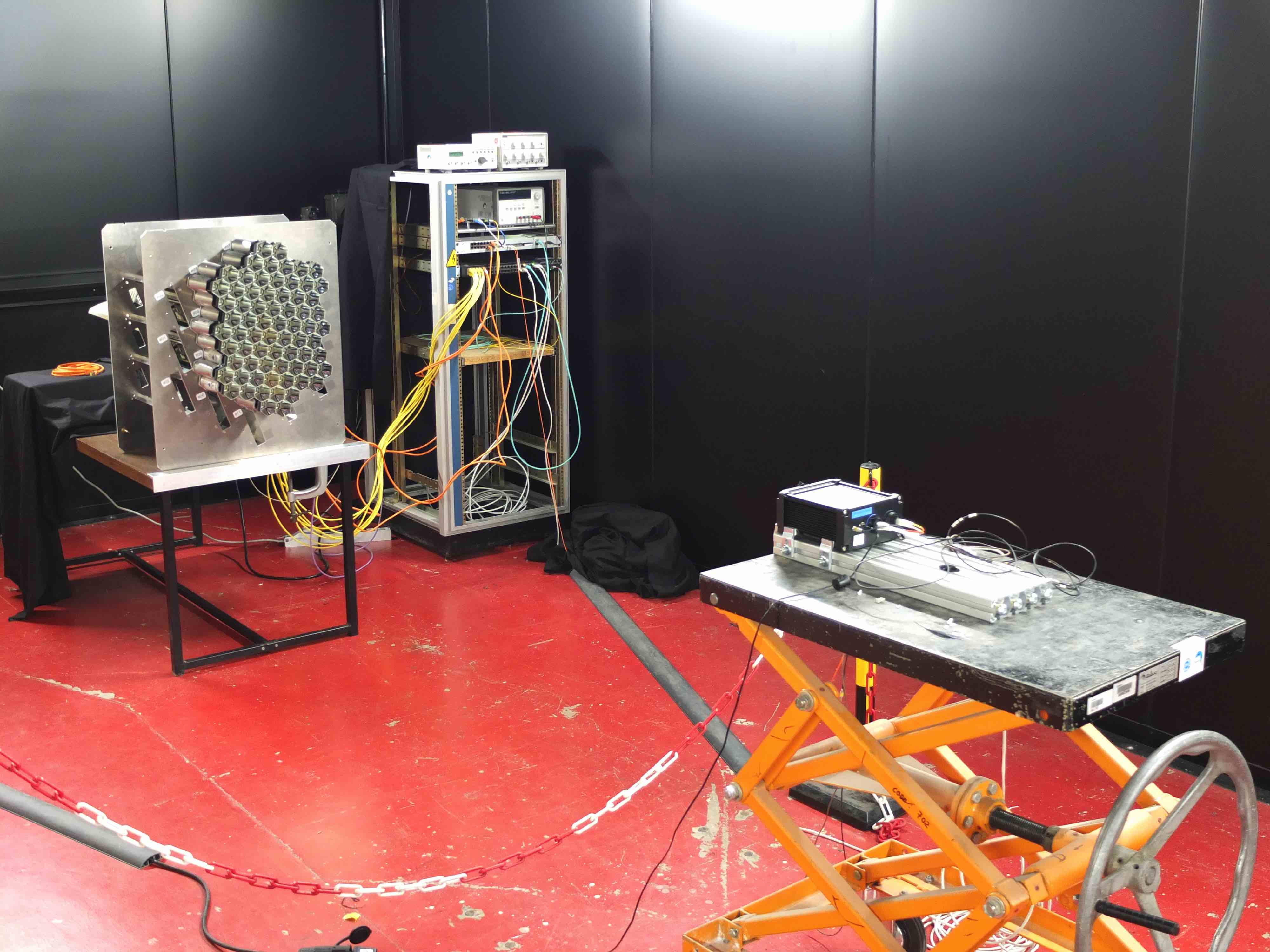}}
  \caption{Left: 19-module mini camera, equipped with {\em digital trigger} backplanes, right: 19-module mini camera in the dark room, equipped with {\em analogue trigger} backplanes. The light source on the right is at a distance of $\sim3$m from the Winston cones. The crate contains the Trigger Interface Board.}
  \label{fig:19module}
\end{figure}

The NectarCAM prototyping strategy is to first test individual components, then mini-cameras with an increasing number of modules. 
The 19-module mini-camera, shown in Fig. \ref{fig:19module}, is aimed at validating trigger options, data acquisition and calibration procedures. 
NectarCAM is evaluating two different camera trigger system designs. The {\em Analogue Trigger} (see Reference \cite{2013ITNS...60.2367T} for details) is based on the analogue summation of the signal of neighbouring pixels, along with summation of L0 signals at camera trigger level (L1). This trigger scheme is known to be very efficient at low energies 
and will be installed on the LSTCAM.  The other possibility (see Reference \cite{Shayduk})  is the {\em Digital trigger} which uses combinations of digitized input at L0. This fully digital trigger scheme is not as efficient for triggering at low energies but has more flexible for changing trigger algorithms. A variant of this trigger design is used by the upgraded H.E.S.S. cameras (see reference \cite{2015arXiv150901232G} for details).

In the analogue trigger scheme, at trigger level L0, pixel signals from a
single module are summed, and the resulting L0 signal distributed to the neighboring modules to perform additional sums. This final sum of neighboring L0 signals is compared against a threshold in a time-window of 3-4 ns to obtain a camera 
trigger decision (L1). In a similar fashion, in the Digital Trigger, the discriminated L0 pixel signals are digitally combined in its own L1 backplane to   produce the camera level trigger decision (L1). 
Once the L1 trigger decision has been locally taken, it can be distributed either to neighbouring modules or to the whole camera. 
There are common components to both the Analogue Trigger and the Digital Trigger design, namely the Trigger Interface Board (management) and the L0 ASIC. 
The L0 ASIC processes analogue signals from the individual pixels in a module. The signal from the pixels are first sent to analogue delay lines, which compensate for the different time shifts introduced by phototubes with different high voltage settings. Two different triggering strategies have been included in the L0 ASIC: the Majority trigger and the Sum trigger. In the Sum trigger concept,
 the pixel signals are clipped and then added together before being sent to the L1 decision trigger.
The Majority trigger concept compares the signal from each pixel to a voltage threshold (also programmable) using a discriminator circuit. Each differential output pair of the discriminator is available as a LVDS digital output of the L0 ASIC. 

Both analogue and digital options 
implement part of their functionalities in dedicated Back-Plane Boards (respectively called Analog Trigger Back-Plane and Digital Trigger Back-Plane), which have the same interface to the front-end board, that enables the testing of both systems with the camera demontrators.
The backplane board is the only electrical interface of the front-end board to the outside world, so it
must deliver all the additional servicing required for normal
operation of the module. Thus, in addition to the trigger functionalities, it provides the front-end
board with Ethernet (for data transfer) and power interfaces,
as well as an IP address.  Finally, the backplane board
receives from the L1 distribution system the synchronization signals required by the front-end electronics.

\begin{figure}[h]
  \centerline{\includegraphics[height=7cm]{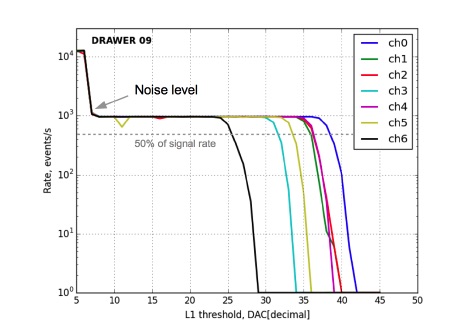}\includegraphics[height=6cm]{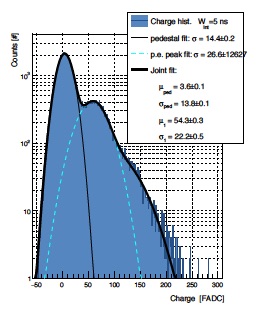}
  }
  \caption{Left: L1 trigger rate scan for 7 channels of a Nectar module. An electronic pulse with a  30 mV amplitude is injected at the input of the front end board. Right: single photoelectron distribution obtained at a 40k photomultiplier gain
  .}
  \label{fig:testbenchresults}
\end{figure}

Trigger options are being validated with electrical signals (left of Fig. \ref{fig:19module})  and with a light source (right of Fig. \ref{fig:19module}).  The light source is installed in a dark room which is large enough to hold an entire camera.
The dark room has a length of $\sim15$m so as to perform flat fielding, pixel linearity response and pixel timing calibrations in a configuration close to the on-site geometry which has the light source located 
on the telescope structure. Both instrument and science calibrations have been performed. The left part of Fig. \ref{fig:testbenchresults} shows a L1 trigger rate scan. The aim is to tune the L1 trigger discriminators 
to obtain similar trigger rates on every pixel of modules. 
The readout chain noise is at the level of $\sim0.2$ photoelectrons.   
The absolute gain calibration of the electronic chain has been performed with single photoelectron distributions (see right part of Fig. \ref{fig:testbenchresults}). On-site calibrations will include single-photoelectron
fits, performed with an XY-staged LED-based flashing screen between the focal plane and the entrance window, as well as muon ring calibrations. 

The camera dead-time is due mostly to the NECTAr chip read-out time and also to a smaller extent to the read-out management (busy signal). The NectarCAM deadtime has been measured with Nectar modules and the full data acquisition and trigger system.  It is under 5\% up to a trigger rate of 7 kHz for a read-out window of 60 ns, and up to 9 kHz for a shorter window of 48 ns.

\begin{figure}[h]
  \centerline{\includegraphics[height=6cm]{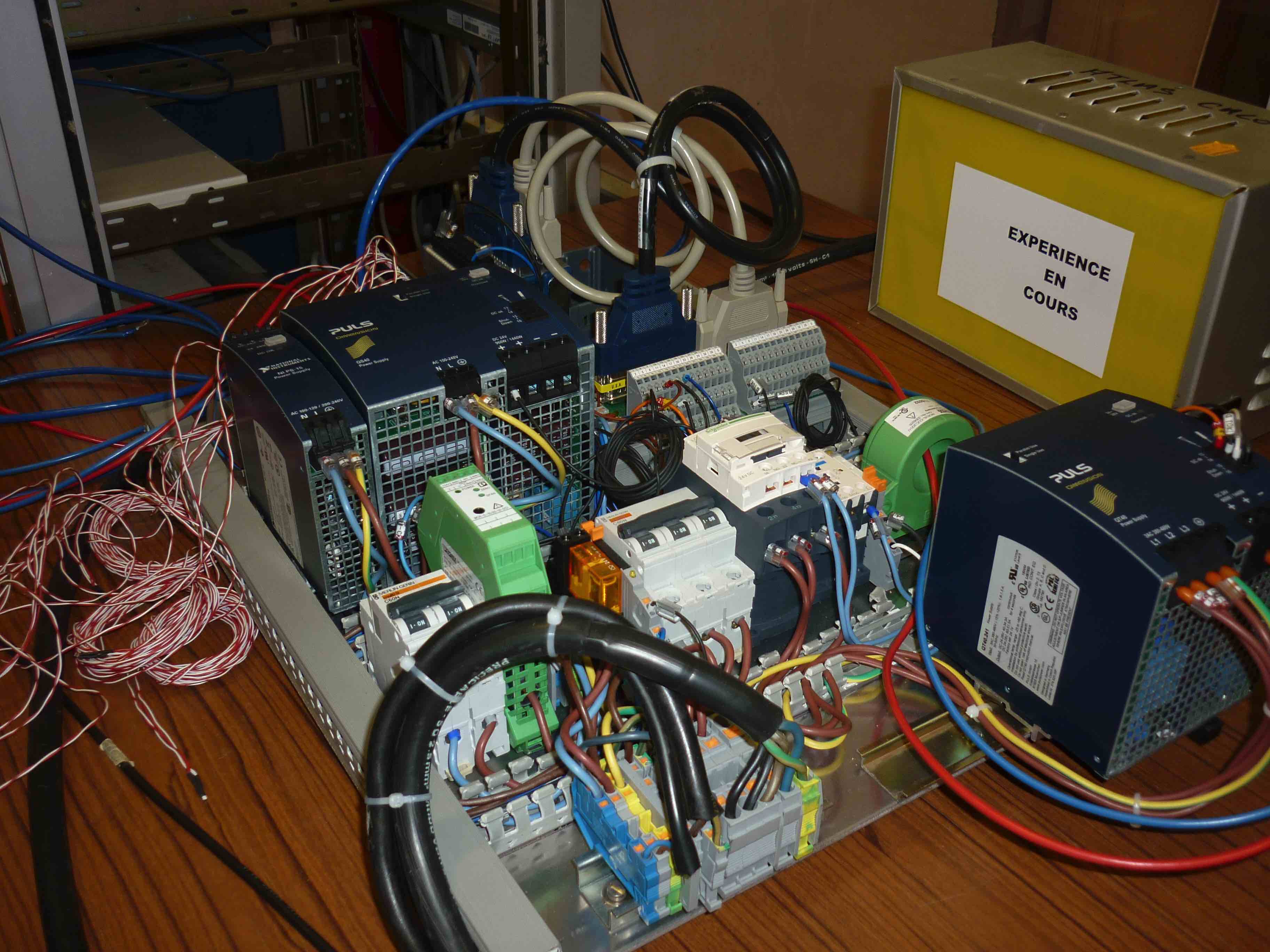}\includegraphics[height=6cm]{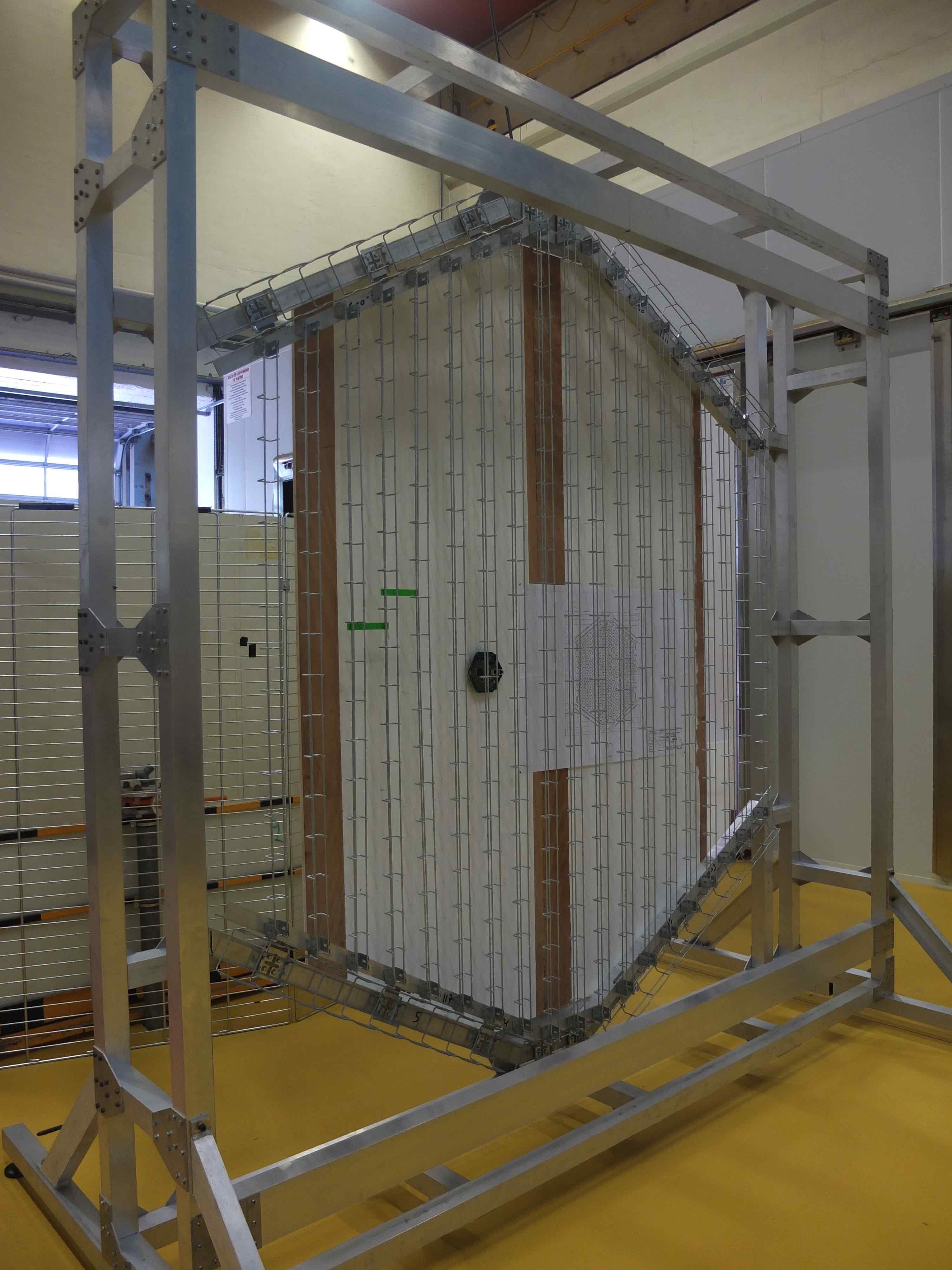}}
  \caption{Embedded camera controller (left) and back-structure prototype (right)}
  \label{fig:ecc}
\end{figure}

\section{Plans for a full-sized camera}
The NectarCAM consortium is now
designing a 70-module (1/4 camera)
partially equipped camera (CANEVAS).
It will include the mechanics, cooling, DAQ, 
and slow-control of a complete
camera. 
As mentioned earlier, all the individual hardware components and most software components have been prototyped. 

The left-hand side of Fig. \ref{fig:ecc} shows the embedded camera controller.  
This device ensures safe conditions for the camera components. It controls the camera subsystems: shutter, low voltage power suplies, temperature and humidity,
monitors the status of the camera and stores the values of state variables in the instrument
database. Finally, it enforces the safety protocols of the camera and takes the relevant
decisions to protect the camera in case of abnormal situations. 
The right-hand side shows a mechanical prototype of the back structure. The center of the structure will hold the cable tray network.  The edges of the structure will hold the 24V power supply units, the trigger interface board which performs the trigger management, the embedded camera controller and the data acquisition switches. 
The partially equipped camera should
be completed in 2017 and could be
mounted on a medium-sized telescope structure
at an astronomical site. Since the
camera is modular, the CANEVAS
camera will be transformed into the
qualification model of NectarCAM as
soon as full funding is available.

\section{ACKNOWLEDGMENTS}
NectarCAM is supported by grants from Agence Nationale de la Recherche, P2IO, OCEVU and OSUG2020 labex, DIM-ACAV from Ile-de-France and by other agencies listed on page: https://portal.cta-observatory.org/Pages/Funding-Agencies.aspx.


\bibliographystyle{aipnum-cp}%
\bibliography{nectarcam}%

\end{document}